\documentclass{article}

\title{On the Cosmological Aspects of Observed High Energy Cosmic Phenomena}
\date{Match 20, 1999}
\author{Anatoli Vankov\thanks{
 Southwest Missouri State University, Department of Physics,
 Astronomy and Materials Science, anv175f@smsu.edu}}

\begin{document}
\maketitle

\centerline{\bf Abstract}

 \small {Super-high energy corpuscular and gamma rays as well as cosmic
 high--power density sources are hard to explain in a galaxy model framework.
 Attempts to include some of those phenomena in the Standard Cosmological
 Model also encounter serious difficulties.  In the present paper an
 alternative cosmological concept is discussed.  There are several features
 in it.  First of all, the whole Universe (Grand Universe) is a multitude
 of typical universes, like ours, evenly made of either matter or antimatter,
 hence, there is no violation of the baryon symmetry on the largest scale.
 Second, high-energy phenomena are the result of matter-antimatter
 annihilation processes in a typical universe evolution.  Finally, the
 Ground Universe is a self-creating due to a balance of annihilation and
 pair creation in the inter-universe infinite space.  This concept and its
 consistence with the major observational data are discussed in detail.}

\subsection*{Introduction}

 As is known, there is presently no physical theory of the universe : the
 Standard Cosmological Model (SCM) seems to fail to explain many Observed
 Universe features because of the singularity problem.  The old question
 why the universe is on average uniform and isotropic is still unanswered.
 The situation is aggravated, as new observational data on high-energy
 processes become available.  In the suggested alternative cosmology the
 relativistic properties of a universe matter outside the Observed Universe
 are revealed.  It gives us a clue for explaining major SCM difficulties. 

\section{CBR as a ``neo-ether''}

 In the SCM the whole Universe, which is the Observed Universe, represents
 a massive absolute reference frame matched with a Cosmic Background
 Radiation (SBR).  The latter is even more perfect absolute reference
 system because it does not have peculiar velocity dispersion.  Actually,
 this issue is similar to the one formulated in the question : why the
 Universe is so well ``tuned'' and ``aligned''.  However, the ``neo-ether''
 issue emphasizes more distinctly a confrontation of the SCM with conventional
 Physics in interpretation of the cosmological observations.  If the Universe
 has such an inherent property as an absolute reference system then special
 and general relativistic theories must be redone.  It is not easy to accept
 a CBR explanation of where the matter goes to in the expanding universe.
 But it is unacceptable to blame the general relativity theory for a decrease
 of a CBR energy density by one expansion factor faster than a total energy
 density of a massive corpuscular matter \cite{[1]}.

 The situation radically changes if one suggests that the Observed Universe
 is not the whole Universe but an ensemble of material objects in the Grand
 Universe.  The latter is a multitude of different ensembles.  Then a CBR as
 ether (an absolute reference frame) and other ``strange'' universe attributes
 become local features of the individual material system.  This idea is put
 in a basis of the suggested alternative cosmological concept.

\section{Antimatter issue}

 Antimatter is apparently present in the Observed Universe in small
 quantities while Physics shows no preference of matter over antimatter.
 If the whole universe is really baryon asymmetric then Physics must be
 deeply revised.  The only way to save a baryon symmetry and a baryon
 charge conservation is again to suggest that the Observed (Home) Universe
 actually is for some reason a matter-made material system.  One may call
 our Home Universe a representative of multitude of typical universes
 evenly made of either matter or antimatter and chaotically dispersed
 in a Grand Universe space.  This is a continuation of the above idea
 of our Home Universe being a material system limited in volume.  Thus,
 resolving the confrontation in baryon symmetry issue we can return to
 the characteristics of the whole Universe (Grand Universe, GU, for short)
 and to the concept of its evolution. 

\section{A Universe without absolute reference frame}

 An absence of the absolute reference system in the GU means a Lorentz
 invariance of a coordinate-momentum distribution function of a GU matter.
 It is known from statistics of a relativistic non-interacting gas \cite{[2]}
 that this function takes a form :
 $$F(x_1,x_2,x_3;p_1,p_2,p_3)dx_1dx_2dx_3dp_1dp_2dp_3 =
  Const\,dx_1dx_2dx_3dp_1dp_2dp_3 \eqno{(1)}$$
 It characterizes a fully chaotic motion.  In this idealized model an
 integral over a momentum distribution (1) diverges.  It shows unlimited
 sky brightness (Olbers paradox).  In practice a bolometer placed in GU
 space should measure a certain average energy density.  It integrates
 radiation (matter flux) over a source distribution, which is a ``last    
 scattering sphere''.  Hence, a real momentum (energy) distribution must
 have a smooth cut--off of upper energies.  This distribution is also a
 Lorentz--invariant because a last scattering sphere does not depend on
 a choice of a reference system.

 In terms of a velocity distribution of relativistic particles a special
 relativity theory gives a formula equivalent to (1), (see, for example,
 \cite{[2]}) :
 $$ F(\beta) d\beta = Const \gamma ^4 \beta ^2 d\beta \eqno{(2)}$$
 for $0<\beta <\beta_{max}$ , where $\beta_{max}$ characterizes a
 cut--off parameter.

 The above relativistic distribution of GU matter is seen unchanged in any
 free reference system.  All forms of matter are supposed to be subjected
 to this law.  Hence, a thought observer can not distinguish between his
 states of relative motion in principle.  Evidently, he can find any
 reference system among multitude of local ones, none of them being absolute.
 The important conclusion is that a GU matter can exist in a state of chaotic
 relativistic motion, which is seen identical for observers in any moving
 free reference system.  In space with a matter in this state any kind of
 an ``absolute'' reference system allowing to detect any effect due to the
 relative motion does not exist.  Now we can develop the alternative
 cosmological concept suggesting that the GU matter is evolving in a
 selfcreating manner being in a baryon symmetric steady state, which is
 found as a Lorentz--invariant (chaotic) relativistic motion.  The GU
 space is thought to be of Minkovski's type.  This is a scene for a cyclic
 evolution of a multitude of typical universes.  Our Home Universe is one
 of them.

\section{On a GU matter theory}

 The Grand Universe is an open system: there is no physical boundary.  This
 is a primary reason for a GU matter being in a stationary state of a
 relativistic motion.  Another important property of the GU system is a
 presence of two complementary entities, that is matter and antimatter.  We
 may describe such a system in terms of a global conservation of sum of two
 kinds of energies : one is ``locked'' in a relativistic mass of matter and
 antimatter, the other is ``released'' in a matter-antimatter annihilation.
 Imagine for a moment that an initial state of the system is ``locked''
 energy.  Than an immediate process of energy release will start in a form of
 explosion (a GU version of ``inflation'' model).  The other extreme case
 would be a state of full matter-antimatter annihilation.  A ``pendulum''
 starts going back through the process of pair creation restoring a
 selfsustained state of a relativistic motion.  There is no energy
 dissipation, and the system acquires an equilibrium state in a form of
 continuing matter-antimatter annihilation and pair creation in parallel.
 Probably, this state may be described in terms of a maximal entropy of
 the GU system.

 It is clear now that we describe the idea of a generalized matter
 transport  theory, which is different from Boltzmann's one first of
 all in above discussed two aspects : openness of the system, and matter
 annihilation/creation.  Besides, we should introduce into this theory
 the basic laws of matter interactions, from gravity mechanics and physics,
 and nuclear physics, in particular.  Boltzmann's statistics of gravitating
 objects is known as a mathematically very complex problem.  Hence, we may
 think about model approximations on different levels.

 The most surprising what was found by author in a simple GU model is a
 statistical separation of matter and antimatter with a formation of an
 hierarchy structure, including evolving ``typical universes'' up to the
 size of our Home Universe.  Evidently, only the smallest chances are
 given to a ``typical universe'' to reach a mature age.

 We expect that a development of a GU matter transport theory will give us
 a basis for a physical theory of Cosmology with a following development of
 specific models of a universe evolution, galaxy formation, cosmological
 nucleonic synthesis and others.  It seems that the theory requires only
 one adjustable parameter : a matter density.  We can not exclude even that
 a more general model may be developed with ``matter density long range
 waves''.  In any case, the purpose of this theory would be solutions of
 the GU transport equation describing both a stationary Lorentz-invariant
 momentum distribution (discussed above) and an ``evolutionary function''
 describing dynamics characteristics of matter structure, mass distribution
 of gravitationally linked systems (like evolving typical universes), in
 particular.

 For further discussing the main topic of this report we need a qualitative
 picture of the GU scenario of our Home Universe evolution.

\section{Our Home Universe evolution}

 Any typical universe should start with some gravitating coagulant.  Its
 growth to an embryonic or a mature state may be thought as the result of
 a probabilistic survival.  Over a lifetime it continually interacts with
 the GU environment, some of the interactions being random ``catastrophic''
 events.  The mass distribution function mentioned must characterize this
 process that is actually fluctuations of the GU matter.  This function
 evidently is a monotonous one rapidly decreasing.  A ``tail'' of the
 function is due to the extremely rare and huge fluctuations what are
 typical universe formations.  A random ``soft'' collision between them
 resulted in either further growth or a major annihilation.  Typical
 universes as they reach a ``mature'' size become unaffected while capturing
 a ``small stuff'' (due to a good statistical averaging).  For survival they
 compete with each other on a similar size level.  In this scenario any
 typical universe is a gravitationally linked relativistic ensemble
 characterized by a momentum distribution similar to one of the GU matter
 but with a lower cut--off parameter.  Eventually a growing universe becomes
 vulnerable to the collision with a quite smaller anti--counterpart.  The
 criterion for this stage may be written roughly in a form:
 $$ G m / R c^2 (\gamma - 1) < 1 \eqno{(3)} $$
 where : $m$ --- mass, $R$ --- mean radius, $\gamma$ --- effective Lorentz
 factor, $G$ --- gravitational constant, $c$ --- speed of light.

 Tracing back to redshifts $z\approx 10$ we can imagine our Home Universe
 being about one order smaller and about three orders denser than at present,
 what gives the criterion (3) a value close to the critical.  If so, a
 mechanism triggering our early Home Universe to decay might be an abrupt
 mass drop due to an accidental collision with a smaller antimatter universe.
 As a result, we observe our present ``expanding'' Universe originated from
 some pre-expansion stage.  Evidently, galaxies (many of them were formed
 at pre-expansion stage) are aligned in a Hubble's flow reflecting an
 initial relativistic distribution of a universe matter.  So, chaos turned
 into an order.

 To complete the rough picture of Our Universe in the GU concept frame we
 should note that stars, galaxies and clusters are inner formations in a
 general GU structure hierarchy. {\it A luminous matter is expected to be
 only a small fraction of a total mass that provides for a self-sustained
 galaxy evolution.}  Hence, a ``dark matter'', which is an ordinary matter,
 should be the dominant part in a typical universe.  If so, it has to be in
 equilibrium  with a thermal radiation.  In our expanding Universe the
 observed 2.7 K CBR bears information on a mean surface temperature of dark
 matter bodies (dust clouds included).  In accordance with the GU concept a
 mass distribution of dark matter is expected to be broad enough what makes
 a universe space quite transparent.  Dark matter naturally participates in
 an ``expansion'' process.  One can easily find that this concept perfectly
 explains all observed CBR features, a temperature decrease inversely
 proportional to the expansion factor, in particular.  In a closed system
 like Our Universe, a thermal radiation performs an adiabatic expansion work
 because of a pressure gradient across the universe volume.  Hence, there is
 no paradox with a ``missing energy''.

 Now we are ready to return to the main topic of this report : to give a
 qualitative explanation of high-energy cosmic corpuscular and gamma rays,
 gamma bursts, quasars and jetting objects in the frame of the suggested
 alternative Cosmology.

\section{Cosmic rays}

 According to the GU concept a space outside typical universes is filled
 with a baryon symmetric matter in relativistic motion characterized by a
 Lorentz--invariant momentum distribution.  The matter includes highly
 energetic gas and all sorts of macroscopic objects.  Any typical universe
 evolves due to the active interaction with this outer GU background.  Our
 Home Universe being also exposed to this background radiation must reveal
 products resulted from a transport of this radiation through the Universe
 medium.  So called primary cosmic rays (its high--energy tail) should be
 actually a secondary radiation from interuniverse sources.  In previous
 author's work \cite{[3]} this cosmic rays transport model is substantiated
 by  numerical assessments.  Cosmic rays researchers long time looked for
 natural  mechanisms of acceleration.  This is one of the old cosmological
 problems.  The GU concept suggests an explanation of this phenomenon,
 paradoxically, as a contrary process of a moderation of primary
 ``inter--universe'' particles (both corpuscular and gamma quants) with much
 higher initial energies than observed.

\section{Star explosions and gamma bursts}

 Gamma bursts discovered not long ago have been so far as enigmatic as cosmic
 rays.  They happen to flash randomly and uniformly in the Universe space.
 Our simple explanation is as follows.  The Universe space in fact contains
 an appreciable amount of antimatter in a form of material objects left from
 the moment of ``Big Bump''; some captured later from the outer space.  As
 mentioned before, a ``dark matter'' partially consists of it.  Some objects
 might be single stars.  Hence, high--energy cosmic gamma rays should have a
 component due to the annihilation processes.  A proper physical model is
 needed to distinguish this component from many other sources.  But part of
 a radiation of an annihilation type is thought to be clearly observable due
 to its pulse character. There must be comparatively rare events of collision
 of matter and antimatter solid objects of big masses.  In particular, the
 whole star may be involved in an annihilation event observed as a gamma
 burst of huge energy.  The data already available on the identification of
 luminous ``disappearing'' objects as gamma burst sources.  This explanation
 is given here as a hypothesis in the GU concept frame for a numerical test.

\section{Quasars}

 We have to explain several quasar features : small size, big power density,
 age comparable with one of the Universe, power variability.  According to
 the GU concept, quasars are ``afterglow'' resulting from the ``Big Bump''
 what was a collision of our early Universe with some smaller typical universe
 made of antimatter (scenario discussed in previous sections).  It must be
 the most catastrophic event in Our Universe evolution.  A huge amount of
 matter has been annihilated with following release of energy in different
 forms of radiation, gamma radiation, in particular.  A quasar ``engine''
 should be a comparably small power--generating ensemble of material objects
 with a high surface--to--volume ratio.  It keeps burning out in a
 non--exploding manner due to a continually incoming antimatter gas stream.
 One may explain its varying power by gas pressure instability.  We think
 that so--called jetting objects have the same physical nature.  A rough
 estimate of the quasar phenomenon in a simplified model \cite{[3]} shows a
 feasibility of matter--antimatter annihilation power generating mechanism.
 Detailed numerical analysis will be made after a development of the whole
 GU concept.

\section{Conclusion}

 Now again the ``hot issues'' in a cosmological field are the old problems of
 the Standard Cosmological Model : is the Universe flat or open, and what is
 a true value of a Hubble's constant.  As often happened in a science
 history, a resolution of a problem could be in finding that a problem is
 incorrectly formulated due to the wrong basic postulate.  In the suggested
 alternative cosmology the answers to the above questions are : our Universe
 is neither open, nor flat, and an exact value of a Hubble's constant does
 not exist at all, for a peculiar velocity field has been initially created
 and further developed in both transversal and radial directions.  The key
 element in a theory of the alternative cosmology must be the idea of a
 self--creation.  A baryon symmetric world in Minkovski's space seems to be
 a perspective concept for a development of a cosmological theory having an
 explanatory and predictive power and being falsifiable.

\subsection*{Acknowledgement}

 I want to express my gratitude to Professor Nikoly Rabotnov for interesting
 discussions of cosmological problems.  I am also thankful to Professor Ryan
 Gieed for his interest and support of my work.

%\subsection*{References}

\end{document}